\PassOptionsToPackage{dvipsnames}{xcolor}
\documentclass[preprint]{aastex61}

\usepackage{amsmath}
\usepackage{capt-of}   
\usepackage{empheq}
\usepackage{xspace}

\newcommand{\ed}{\textcolor{black}}
\newcommand{\edd}{\textcolor{black}}
\newcommand{\rf}{\textcolor{black}}

\newcommand{\rff}[1]{{\color{black}{#1}}}
\newcommand{\rfGA}[1]{{\color{black}{#1}}}
\newcommand{\rfga}[1]{{\color{black}{#1}}}

\newcommand{\ang}{\ensuremath{\rm \AA}\xspace}
\newcommand{\AV}{\ensuremath{A_V}\xspace}

\newcommand{\avgN}{\ensuremath{\langle N \rangle}\xspace}
\newcommand{\avgAV}{\ensuremath{\langle A_V \rangle}\xspace}
\newcommand{\avgI}{\ensuremath{\langle I_{100} \rangle}\xspace}

\newcommand{\firstdiam}{\ensuremath{160}\xspace}
\newcommand{\lastdiam}{\ensuremath{430}\xspace}

\newcommand{\HST}{\emph{HST}\xspace}

\newcommand{\micro}{\ensuremath{\micron}\xspace}
\newcommand{\sncu}{SN~2012cu\xspace}
\newcommand{\snfe}{SN~2011fe\xspace}
\newcommand{\snJ}{SN~2014J\xspace}
\newcommand{\snX}{SN~2006X\xspace}

\newcommand{\lmax}{\ensuremath{\lambda_{max}}\xspace}
\newcommand{\RV}{\ensuremath{R_V}\xspace}
\newcommand{\EBV}{\ensuremath{E(B-V)}\xspace}

\received{August 15, 2017}
\revised{September 26, 2017}
\revised{October 12, 2017}
\accepted{October 13, 2017}
\submitjournal{ApJ}

\shorttitle{Non-uniform ISM}
\shortauthors{Huang et al.}

\begin{document}
\title{On the Time Variation of Dust Extinction and Gas Absorption for Type~Ia Supernovae Observed Through Non-uniform Interstellar Medium}


\author[0000-0001-8156-0330]{X.~Huang}
\affiliation{Department of Physics and Astronomy, University of San Francisco, San Francisco, CA 94117-1080}

\author{G~Aldering}
\affiliation{Physics Division, Lawrence Berkeley National Laboratory, 1 Cyclotron Road, Berkeley, CA, 94720}

\author{M.~Biederman}
\affiliation{Department of Physics and Astronomy, University of San Francisco, San Francisco, CA 94117-1080}

\author{B.~Herger}
\affiliation{Department of Physics and Astronomy, University of San Francisco, San Francisco, CA 94117-1080}

\begin{abstract}
For Type Ia supernovae (SNe Ia) observed through a non-uniform interstellar medium (ISM) in its host galaxy, we investigate whether 
the non-uniformity can cause observable time variations in dust extinction and in gas absorption due to the expansion of the SN photosphere with time.  We show that, owing to the steep spectral index of the ISM density power spectrum, sizable density fluctuation amplitudes at the length scale of typical ISM structures ($\gtrsim \text{ 10 pc}$) will translate to much smaller fluctuations on the scales of a SN photosphere.
Therefore the typical amplitude of time variation due to non-uniform ISM, of absorption equivalent widths and of extinction, would be small.  As a result, we conclude that non-uniform ISM density should not impact cosmology measurements based on SNe Ia.  We apply our predictions based on the ISM density power law power spectrum to the observations of two highly reddened SNe Ia, \sncu and \snJ. 

\end{abstract}
\keywords{ISM --- supernovae: general, dust, extinction, ISM: structure, ISM: general, cosmology: miscellaneous 
}

\section{Introduction}
\label{sec:intro}
One of the remaining systemic uncertainties for Type Ia Supernovae (SNe Ia) as a distance indicator is dust extinction.  From measurements of wavelength-dependent extinction and polarimetry, the extragalactic dust obscuring light from SNe Ia in a number of cases appears to have different properties compared with dust in the interstellar medium (ISM) of the Milky Way (MW): both the total-to-selective extinction ratio, \RV, and the Serskowsi polarization parameter \lmax \citep{serkowski75a} are often found to be significantly lower
\setcitestyle{notesep={; }}\citep[e.g.,][but see \citealt{amanullah15a} and \citealt{huang17a} for an example of high extinction but MW-like \RV]{phillips13a, amanullah14a, patat15a, zelaya17a}.  
\setcitestyle{notesep={, }}

To account for the ``peculiar" dust properties for SNe Ia, it has been suggested that scattered light by dust in a circumstellar medium (CSM) associated with an SN Ia  
may provide an explanation for both the low values of \RV \citep{wang05a, patat06a, goobar08a} and \lmax \citep{patat15a, hoang17a}.  Observationally, CSM indeed seems to be present for at least some SNe Ia, as evidenced by the detection of SN-CSM interactions \citep[e.g,][]{hamuy2003a, aldering2006a, dilday2012a}, blueshifted and time-varying \ion{Na}{1} or \ion{K}{1} absorption features \citep[e.g.,][]{patat07a,  simon09a, fox13a, sternberg14a, graham15a}, and the imbalance between the blue- and redshifted \ion{Na}{1} absorption components \citep{sternberg2011a, foley2012a, maguire2013a, phillips13a}.  On the other hand, stringent limits on CSM density has been placed on two of the nearest SNe Ia, \snfe and \snJ \citep[e.g.,][]{chomiuk12a, perez-torres14a, margutti14a, brown15a,  harris16a, johansson17a}. 
Indeed, for most SNe Ia, observational evidence points in the direction of interstellar dust being at least  the dominant cause for extinction, including those that are highly reddened \citep[e.g.,][]{patat07a, chotard11a, phillips13a, 
patat15a, maeda16a, zelaya17a, huang17a, bulla2017a}.  
The low values of \RV and \lmax for some SNe Ia may then be the result of the dust grain size distribution in some of the host galaxies being different from the MW and/or possibly having been altered by SN radiation \citep[e.g.,][]{hoang17a}.  In short, the nature and properties of extragalactic dust toward SNe Ia have been a long standing puzzle, and a deeper understanding is important for cosmology, SN progenitor \citep[for a recent review, see, e.g.,][]{maoz2014a}, and ISM studies.


One possible approach that can help unravel this puzzle is to explore whether or how the extinction and the EWs of absorption features for SNe Ia may vary with time.  For CSM, \citet{wang05a} and \citet{brown15a} showed that by including multiply scattered photons the visual extinction, \AV, would decrease with time, owing to 
the delayed arrival of short-wavelength photons.  To-date this has not been observed \rff{for dust in the CSM \rfGA{(though it may have been detected for dust in the ISM \citep{bulla2017a})}}.  
The EWs for some of the absorption features are also expected, and actually observed, to change over time due to variable CSM ionization conditions induced by SN radiation \citep[e.g.,][]{patat07a}.  However,
the EWs are not expected to vary in the same way as \AV.  In fact, for some absorption features, no variation is expected
\citep{patat07a, graham15a}.  

For ISM, \citet[][P10]{patat10a} and \citet[][F13]{foerster13a} simulated column density maps following a power law power spectrum and argued that due to the non-uniformity of the column density and the expansion of the projected SN photosphere, the \ion{Na}{1} absorption EW and extinction, respectively, would likely vary with time.  The effects were forecast to be detectable for the EW and significant for extinction.  In contrast to the CSM scenario, if time variations of both are detected for a SN Ia, they should be correlated.

In this paper, we re-examine the effects of non-uniform ISM density on extinction and absorption EWs, for two purposes.  First, to determine whether the observation of time variation of extinction and absorption EWs can be used to distinguish the two major extragalactic dust scenarios for SNe Ia.
Second, for SN cosmology, it is important to determine the magnitude of time variation for extinction due to interstellar dust with non-uniform column density.  This is especially true as the sum of the observational evidence so far suggests that, though the extragalactic dust properties may be peculiar (i.e., not MW-like) for SNe Ia, the extinction is likely caused predominantly, if not exclusively, by dust in the host galaxy ISM and not CSM associated with SNe.  \rff{Our main departure from P10 and F13 concerns the appropriate level of ISM variation on the scale of a SN Ia photosphere at a given average column density.}

This paper is organized as follows.  In \S~\ref{sec:normalization}, using observations of MW dust emission, we present a way to determine the normalization for the typical ISM column density fluctuations on the scale of SN photospheres at a given column density.
In \S~\ref{sec:observations}, we compare the predictions based on this normalization with the observations of two SNe Ia with high extinction, \sncu and \snJ.  \rff{We also discuss \snX, \rfga{using it to contrast the observational signatures of non-uniform ISM and a light echo.}}  We conclude in \S~\ref{sec:conclusion}.  Throughout this paper, we assume constant dust-to-gas ratio \citep[e.g.,][]{padoan97a, miville-deschenes07a}.  For simplicity, all SN phases are taken relative to $B$-band maximum.

\section{ISM Density Power Spectrum Normalization}
\label{sec:normalization}


ISM column density variations on scales from $\gtrsim 10$~pc to $\sim 500$~AU have been shown to follow a power law power spectrum, having a spectral index $\gamma$ typically between $-2.5$ and $-3.6$ \citep[e.g.,][]{gautier92a, schlegel98a, deshpande00a, padoan06b, miville-deschenes07a, royn10a}.  On smaller scales, column density variations down to $\lesssim 10$~AU have been observed \citep[e.g.,][]{dieter76a, diamond89a, heiles97a, weisberg07a, lazio09a, stanimirovic10a}.  The same kind of power law behavior is found for the small scale fluctuations \citep[e.g.,][]{gibson07a, royn12a}.  \citet[][D00]{deshpande00b} used 1D simulations to demonstrate that a scale-free power law power spectrum with $\gamma = -2.75$ extending from a few pc down to AU scales can explain the frequency of small scale structures in \ion{H}{1}.  
More recently, \citet{dutta14a} found for ISM density fluctuations on $\sim 10$~AU scales that the spectral index is the same as that for pc scale fluctuations and the upper limit for the amplitude is consistent with the pc scale fluctuation amplitude. This bolsters the suggestion made by D00 that ISM density fluctuations on scales from a few pc down to $\lesssim 10$~AU can be described by a scale-free, statistically isotropic and homogeneous power law power spectrum.

\subsection{Summary of Relevant Results in P10 and F13}\label{sec: summary-P10-F13}

P10 pointed out that if there are ISM structures on such small scales in the MW then SNe Ia can be used to probe small-scale fluctuations in the host galaxy ISM.  
The SN photospheres have diameters of up to $\sim 800$~AU during the photospheric phase (within 100 days after explosion (P10)).
As the photosphere \edd{(and its projection, the photodisk)} of a SN Ia expands, \edd{its light} is essentially sampling the average enclosed ISM column density.  Thus, for example, the \ion{Na}{1} absorption feature may vary with time (this would apply to the absorption of other atomic or molecular species if their column density is in the right range; see Section~\ref{sec:sn14J}).
Following the suggestion in D00 that the power law power spectrum extends down to $\lesssim 10$~AU scales, P10 explored this possibility by simulating ISM column density maps with a width of 1024~AU and $\gamma = -2.8$.
P10 considered the cases where the peak-to-peak fluctuation amplitudes were $\Delta N = \avgN$ and 2\avgN, and showed in each case that there could be observable variations of EW(\ion{Na}{1}) between phases of $-$10 and +50 days, depending on the mean density, \avgN.  During this period, the photodisk increases from approximately 120~AU to 640~AU across (P10).

In a similar way, F13 simulated dust column density maps.  Compared with P10, they used $\gamma = -2.75$ (slightly shallower), a width of 2048~AU (twice as large), 
and a peak-to-peak fluctuation amplitude of $\Delta N = 2\avgN$ (similar).  They found that the SN extinction, which is proportional to the mean column density over the photodisk, would likely vary with time by a substantial amount over a period spanning 140 days from the time of explosion. 
Their Figure~8 showed especially dramatic variations within 60 days after explosion, corresponding to a phase of approximately 43 days \citep[e.g.,][]{hayden10a}.  \ed{This is the period when most light curve measurements for cosmology applications are made.}

\ed{Neither P10 or F13 made clear how the fluctuation amplitudes for their simulated ISM density maps} were chosen.
In the next subsection we will set the normalization for the power spectrum based on MW dust emission observations and this will in turn \edd{provide} the typical amplitude of fluctuation on the scale of a SN photodisk.

\subsection{Normalization for ISM Fluctuations}\label{sec:norm}
Given a power law power spectrum for a 2D density field, $P(k) \sim k^\gamma$, one can compute the structure function \citep{lee75a}, 
$S(d)$, which is the variance for two points separated by a distance $d$: 

\begin{equation}
S(d) = \sigma^2(d) \sim d^{-\gamma - 2}
\end{equation}

\noindent
This is approximately the same as the variance within a disk with diameter $d$ \citep[see, e.g.,][]{brunt02a}.   This result can be obtained by exploiting the approximate equivalence between a circular tophat window function with diameter $d$ in configuration space, corresponding to the photodisk, and a circular tophat window function with a diameter $\sim 1/d$ in $k$-space, and then integrating $P(k)$ outside of this window function.

\citet[][MD07]{miville-deschenes07a} re-analyzed the IRAS data for 55\% of the sky. 
From the intensity map of the 100~\micro emission from cirrus clouds, they found a power law for the power spectrum at the pc scale (see below concerning the conversion from angular to physical scales) with $\gamma = -2.9 \pm 0.2$.  Using the relationship between the power spectrum and the structure function, they provided the following normalization for \rff{the RMS of the 100 \micro emission in} the regime $\langle I_{100} \rangle > 10 \text{ MJy/sr}$ \footnote{Note the power in Equation~(18) of MD07 should be 1.55 instead of 1.5.  This follows from Equation~(8) in MD07.  Otherwise there would be a 10\% mismatch for the normalization at the \avgI = 10~MJy/sr boundary.},


\begin{equation}
\rf{\sigma_{I_{100}}(\theta)} = 0.1 \avgI^{1.55} \left( \frac{\theta}{12.5^\circ} \right)^{-\gamma/2 - 1}
\label{eqn:angular-norm}
\end{equation}

\noindent 
where $I_{100}$ is the intensity of the 100 \micro emission and $\theta$ the angular scale of interest.  

\rff{\subsubsection{Normalization for Extinction Fluctuations}\label{sec:norm-AV}}

The 100 \micro emission can be converted to extinction using the precepts of \citet{schlegel98a}

\[
\avgAV \approx 0.050\langle I_{100} \rangle
\]
 
\noindent
Thus the condition $\langle I_{100} \rangle > 10 \text{ MJy/sr}$ corresponds to $\avgAV > 0.5$~mag.  


To place these results on the scale of a SN photosphere, we also need to convert the angular reference scale used in MD07 ($12.5^\circ$) to a physical scale.  The distances to the cirrus clouds found in the literature are typically between 100~pc and 750~pc \citep{benjamin96a, grant99a, szomoru99a}, with one group reporting a distance of 1.8~kpc \citep{jackson02a}.  These distances are consistent with \edd{a minimum distance set by} the ``local cavity" reported by \citet[][and references therein]{welsh10a} --- roughly within a 80~pc radius around the solar system, there is little \ion{Na}{1}, often used as a tracer for dust.  Thus conservatively we use a distance of 80~pc, which converts $12.5^\circ$ to approximately 18~pc; using larger distances would further suppress the implied density fluctuations on the scale of a SN photosphere.  In addition, the fact that MD07 dealt with fixed angular scale meant variance on larger and larger physical scales were added in the integral for the total variance along the line of sight, whereas we need a fixed physical reference scale.  One can work out the geometric correction factor, $1/\eta = \sqrt{-\gamma - 1}(h/H)^{-\gamma/2 - 1}$, where $h$ and $H$ are the lower and upper limits of the integral for the total variance.  Taking the lower limit of $h = 80$~pc, $\eta$ is typically between 1.5 and 3, depending on whether one sets the upper limit $H$ to be 750~pc or 1.8 kpc, and depending on the value of $\gamma$.

We can now rewrite Equation~\ref{eqn:angular-norm} as

\begin{equation}
\sigma_{\AV}(d) = \left(\frac{0.52}{\eta} \right) \langle \AV \rangle^{1.55} \left( \frac{d}{18 \text{ pc}} \right)^{-\gamma/2 - 1}, \hspace{2 cm} \avgAV > 0.5\text{ mag}.
\label{eqn:lin-norm-hiAV}
\end{equation}


\noindent
To get a better sense of the expected fluctuation amplitude for \AV, we will use 800~AU as the reference scale, which is the extent of the photodisk at the end of the photospheric phase.  Assuming a typical value of $\gamma = -2.9$, we obtain, 
\begin{equation}
\sigma_{\AV}(d) = \left(\frac{0.012}{\eta} \right) \langle \AV \rangle^{1.55} \left( \frac{d}{800 \text{ AU}} \right)^{0.45}, \hspace{2 cm} \avgAV > 0.5\text{ mag}.
\label{eqn:lin-norm-SN-hiAV}
\end{equation}

\noindent
\ed{The comparison between Equations~\ref{eqn:lin-norm-hiAV} and \ref{eqn:lin-norm-SN-hiAV} shows that a key factor in this \edd{conversion} is $\left( \frac{\text{800 AU}}{\text{18 pc}} \right)^{-\gamma/2 -1} = 0.022$ for $\gamma = -2.9$}.

The normalization for $\langle I_{100} \rangle < 10 \text{ MJy/sr}$ has a different dependence on $\langle I_{100} \rangle$ (MD07).  Most SNe Ia used as cosmological probes have $\AV < 0.5$~mag, corresponding to this case.  Analogous to Equations~\ref{eqn:lin-norm-hiAV} and~\ref{eqn:lin-norm-SN-hiAV}, then we have, 

\begin{equation}
\sigma_{\AV}(d) = \left( \frac{0.35}{\eta} \right) \langle \AV \rangle  \left( \frac{d}{\text{18 pc}} \right)^{-\gamma/2 - 1}, \hspace{2 cm} \avgAV < 0.5\text{ mag}
\label{eqn:lin-norm-loAV}
\end{equation}

\begin{equation}
\sigma_{\AV}(d) = \left( \frac{0.0078}{\eta} \right) \langle \AV \rangle  \left( \frac{d}{800 \text{ AU}} \right)^{0.45}, \hspace{2 cm} \avgAV < 0.5\text{ mag}.
\label{eqn:lin-norm-SN-loAV}
\end{equation}

\noindent
Just as with Equation~\ref{eqn:lin-norm-SN-hiAV}, in the conversion from using 18~pc as the reference length to 800~AU for Equation~\ref{eqn:lin-norm-SN-loAV}, for convenience we set $\gamma = -2.9$.  If we assume that this $\sigma_{\AV}(d)$-$\langle \AV \rangle$ relationship for the MW applies to SN host galaxies, then Equations~\ref{eqn:lin-norm-SN-hiAV} and \ref{eqn:lin-norm-SN-loAV} make clear that during the photospheric phase the extinction due to non-uniform interstellar dust is not expected to vary much for a SN Ia.  

\ed{Finally, it is instructive to see how the extinction fluctuation RMS depends on the phase.  Employing the equation for the diameter of the photodisk $t$ days after explosion given by P10,
in terms of the phase $T$ \citep[= $t$ + 17.4; e.g.,][]{hayden10a}, for $\gamma = -2.9$, we can write:}

\begin{equation}
\sigma_{A_V} (T) \sim \left( \frac{ \left(7.0 + 10.6\, e^{-\edd{(T - 17.4)} /36.5} \right) \, \edd{(T - 17.4)}} {800\text{ AU}} \right)^{0.45}
\end{equation}

\rff{
\subsubsection{Normalization for Column Density Fluctuations}\label{sec:norm-N}
To find the typical normalization for the column density fluctuations, we follow similar steps as in Section~\ref{sec:norm-AV}.  From \citet{lagache2000a}, we obtain the conversion from \avgI to $N$(\ion{H}{1}) and given the ratio of $N$(\ion{Na}{1})/$N$(\ion{H}{1}) is approximately constant for the MW \citep[e.g.,][]{murga2015a}, we find

\begin{equation}
N(\text{\ion{Na}{1}}) = 0.5 \times 10^{12} \avgI\,  \rm{cm^{-2}}
\label{eqn:N-Na-conversion}
\end{equation}

\noindent
From this relation, 
we obtain the RMS of the column density fluctuation,

\begin{subequations}
    \begin{empheq}[left={\sigma_N(d)= \left( \dfrac{d}{18 \text{ pc}} \right)^{-\gamma/2 - 1} \rm{cm}^{-2}\empheqlbrace\,}]{align}
      &  \left(\frac{1.5\times 10^{11}}{\eta} \right) \left(\frac{\avgN}{10^{12}} \right)^{1.55} \mkern-36mu 
      &, \quad & \avgN > 5 \times 10^{12}\, \rm{cm^{-2}} 
      \label{eqn:N-norm-hi}\\
      & \left(\frac{3.5 \times 10^{11}}{\eta} \right) \left(\frac{\avgN}{10^{12}} \right) \mkern-36mu 
      &, \quad &  \text{otherwise}
      \label{eqn:N-norm-lo}      
\end{empheq}
\label{eqn:N-norm}
\end{subequations}

\noindent

Using a reference length of 800~AU (convenient for SN), and the typical value of $\gamma = -2.9$, the RMS of the \ion{Na}{1} column density is given by,

\begin{subequations}
    \begin{empheq}[left={\sigma_N(d)= \left( \dfrac{d}{800 \text{ AU}} \right)^{0.45} \rm{cm}^{-2}\empheqlbrace\,}]{align}
      &  \left(\frac{3.3\times 10^{9}}{\eta} \right) \left(\frac{\avgN}{10^{12}} \right)^{1.55} \mkern-36mu 
      &, \quad & \avgN > 5 \times 10^{12}\, \rm{cm^{-2}} 
      \label{eqn:N-norm-SN-hi}\\
      & \left(\frac{7.8 \times 10^{9}}{\eta} \right) \left(\frac{\avgN}{10^{12}} \right) \mkern-36mu 
      &, \quad & \text{otherwise}
      \label{eqn:N-norm-SN-lo}      
\end{empheq}
\label{eqn:N-norm-SN}
\end{subequations}

\noindent
which makes clear the typical amplitude of fluctuations around \avgN for SNe Ia.}

\section{Comparison with Observations}\label{sec:observations}

In this section we compare the predictions from the previous section with the observations of two well-observed, highly reddened SNe Ia, \sncu and \snJ.  \rfGA{We also discuss \snX, a third highly reddened SN~Ia, and contrast our predictions with the effects of a light echo.}


\subsection{SN~2012cu}\label{sec:sn12cu}

\sncu has one of the highest extinction values ($\AV \approx 2.9$~mag) for a SN Ia \citep{amanullah15a, huang17a}.   Several lines of observational evidence point to dust in the ISM as the dominant, and possibly the only, source for the extinction \citep{huang17a}.

\citet{huang17a} found \snfe and \sncu to be good spectroscopic twins \citep[see][]{fakhouri15a}.  Their observations were also closely matched in phase for 10 epochs between $-6.8$ and +23.2 days.
Using \snfe as the template, the extinction was determined to be $\AV = 2.944$~mag \ed{with a RMS of $0.043 \pm 0.010$~mag.}
This RMS around the time-averaged \AV across the 10 phases likely has some contribution from the fact that
\sncu and \snfe are not perfect twins\edd{, as discussed in \citet{huang17a}}.  
Non-uniform interstellar dust, however, could also contribute to the apparent variation.  Below we explore the implication of this possibility.

Between the phases of $ -6.8$ and +23.2 days, the photodisk has a diameter between \firstdiam and \lastdiam~AU (P10).  To calculate the expected RMS for the fluctuation within a disk of \lastdiam~AU, $\sigma_{\AV}(\text{\lastdiam AU})$, we will set $d = \lastdiam$~AU and $\langle \AV \rangle = 2.944$~mag, and assume $\eta = 2$, in Equation~\ref{eqn:lin-norm-hiAV}.  The results are  summarized in Table~\ref{tab:AV-RMS}.  For completeness we also show \ed{$\sigma_{\AV}(\text{240 AU})$ (at maximum)} and $\sigma_{\AV}(\text{800 AU})$ (corresponding to a phase of approximately 82.6 days, the end of the photospheric phase).


\begin{deluxetable}{lccc}
\tablewidth{0pt}
\tabletypesize{\scriptsize}
\tablecaption{
\label{tab:AV-RMS} Predicted RMS within the Photodisk 
for \sncu at Representative Phases \\\hspace{\textwidth}} 
\tablehead{
\colhead{\ed{Phase}}  &
\colhead{\ed{0 days (at max)}} &
\colhead{\ed{+23.2 days}} & 
\colhead{\ed{+82.6 days}} \\  
\hline
\colhead{$\gamma$}  &
\colhead{$ \sigma_{\AV}(\text{240 AU}) $}  & 
\colhead{$ \sigma_{\AV}(\text{\lastdiam AU}) $[}  & 
\colhead{$ \sigma_{\AV}(\text{800 AU}) $}\\ 
}
\startdata
$-2.9$ (typical) & 0.018 & 0.024  & 0.033    \\
$-2.5$ (lowest)  & 0.13  & 0.15   & 0.17   \\  
\enddata
\end{deluxetable}

\ed{The predicted $\sigma_{\AV}(\text{\lastdiam AU})$ is not quite the same as what was observed, which is the RMS of the observed \AV over time.  The variation across the phases are correlated for a given SN because the SN photodisk at any given phase encloses a smaller photodisk at an earelier observed phase.  However,} the value for this quantity sets the upper limit for how much the extinction \textit{can} vary  across phases up to 23.2 days.


Two things are made clear in Table~\ref{tab:AV-RMS}: 1) even for the shallowest value of $\gamma = -2.5$, the variation is much smaller than what is assumed in 
F13 (where the peak-to-peak amplitude was set to be $\Delta N = 2\avgN$ \ed{on the scale of 2048 AU, or equivalently, $\Delta \AV = 2 \avgAV$, and corresponds to a RMS of $\sigma_{\AV} \approx 14\% \times \avgAV$ on the scale of 430~AU});
2) the small observed RMS around the time-averaged \AV for \sncu is consistent with the expectation for a scale-free power law power spectrum, even though it may be difficult to separate exactly the contributions from non-uniform dust vs. small differences in the intrinsic spectra
between \sncu and \snfe.

Note that Table~\ref{tab:AV-RMS} provides the upper limits for the contribution from non-uniform interstellar dust to the RMS for the observed \AV across phases for \sncu, even if observations earlier than $ -6.8$ days 
for \sncu were included.
                
Finally, for $\AV < 0.5$~mag (relevant for SNe Ia typically used in a cosmological study), the RMS fluctuation from Equation~\ref{eqn:lin-norm-loAV} is typically on the order of 0.01~mag or below, even for the shallowest value of $\gamma = -2.5$, and even if the phase coverage spans the entire photospheric phase ($\sim 100$~days after explosion).

\subsection{SN~2014J}\label{sec:sn14J}

\snJ has an extinction of $\AV = 1.9$ \citep[e.g.,][]{amanullah14a, foley14a, gao15a}.  Despite having a low \RV of 1.4 -- 1.7, here too, many lines of evidence point to dust in the ISM as the dominant, and likely the only, source for the extinction \setcitestyle{notesep={; }} \citep[][\ed{also see Appendix}]{kawabata14a, margutti14a, perez-torres14a, patat15a, brown15a, maeda16a, johansson17a, bulla2017a}.  


\subsubsection{ISM Absorption Variations}\label{sec:sn14J-EW}
P10 made detailed, quantitative predictions for the amplitudes of EW(\ion{Na}{1}) variation by simulating ISM maps for a width of 1024~AU and a spectral index of $\gamma = -2.8$.  They considered different combinations of average column density values ($\langle N \rangle = 10^{11}, 10^{12}, \text{ and } 10^{13}\, \mathrm{cm}^{-2}$), peak-to-peak fluctuation amplitudes ($\Delta N = \avgN \text{ and } 2\avgN$), and Doppler parameter values ($b = 1 \text{ and } 5 \text{ km s}^{-1}$).  The results are tabulated in their Table~2.  They pointed out that for $N > 10^{13} \, \mathrm{cm}^{-2}$, the absorption will be saturated and that as a result, in this case, the EW will only weakly vary with column density fluctuations.  
The \ed{largest EW variation is achieved for} the average column density of $\sim 10^{12}\, \mathrm{cm}^{-2}$.  
For the same column density fluctuation amplitude, a higher value of $b$ would result in greater variation in EW.  

Based on the predictions of P10, it was already clear that attempting to detect small scale density fluctuations of the foreground ISM in the host galaxy by measuring EW variations would be challenging.  \ed{Even with an optimal parameter combination, 
the RMS of the EW variation is $\sim$~a few m\AA\  (see their Table~2).}



\rff{To calculate the column density fluctuation RMS based on the normalization determined in this paper,} we set $d = 640$~AU, the maximum photosdisk extent in P10 at a phase of +50 days, and $\avgN = 10^{12} \text{ cm}^{-2}$, with $\gamma = -2.8$ (same as used in P10) and $\eta = 2$, \rff{and obtain $\sigma_N (640 \text{ AU}) = 5.5 \times 10^{9}\, \rm{cm}^{-2}$ from Equation~\ref{eqn:N-norm-lo}.}  This is much lower than the RMS values used by P10.  For the same mean column density, the level of EW variation is approximately proportional to the fluctuation RMS of the column density (P10).  Thus even under the optimal combination of parameters in P10, the RMS of EW variation would be $\lesssim 0.1$ m\AA.

The comparison between the density fluctuation levels chosen by P10 and F13 and our predictions (using $\gamma = -2.9$) are presented in Figure~\ref{fig:AV-Na-sigma}.


\ed{EW(\ion{Na}{1}) would not be very responsive to column density variations if the \ion{Na}{1} absorption is saturated.  For a SN with multiple non-overlapping absorbing components, among which at least some have an average column density in the neighborhood of $10^{12} \, \mathrm{cm}^{-2}$, the EW fluctuation RMS due to the different components will add in quadrature.  
Thus for the same high total column density, a SN with the optimal component distribution would have the highest variation in EW(\ion{Na}{1}).}  \snJ is close to such a case.
But even in this case, the total RMS for EW variation would still be very small (our rough estimate is $\lesssim 0.2$~m\AA).  
In general, for a multi-component absorption system, if the $b$ values are small, $\Delta$EW would be negligible for each component.  On the other hand, if the $b$'s are large, 
fewer \ed{non-overlapping, unsaturated} components can fit into the available ``velocity space" of a galaxy (on the order of hundreds of km/s).  Thus it is difficult to see a scenario in which the total RMS would rise to even 1~m\ang for SNe Ia due to ISM with non-uniform density.

For the \ion{Na}{1} absorption system, \edd{\citet{foley14a}}, \citet{ritchey15a}, and \citet{graham15a} showed that there were no detectable EW variation between phases $-11$ to +47 days.
\citet{maeda16a} determined the \ion{Na}{1} absorption for \snJ to be \textit{exclusively} in the ISM.
They did not see variation in the absorbing systems toward the SN out to a phase of +255.1 days (or within a pencil beam of $\sim 0.02$~pc).
Furthermore they presented evidence for ISM column density variation on the scale of 20~pc by observing the diffuse light of the host galaxy (M82) around the SN. 
\edd{These findings} are consistent with the predictions of this paper.

\noindent%
\begin{minipage}{\linewidth}
\makebox[\linewidth]{
  \includegraphics[keepaspectratio=true,scale=0.6]{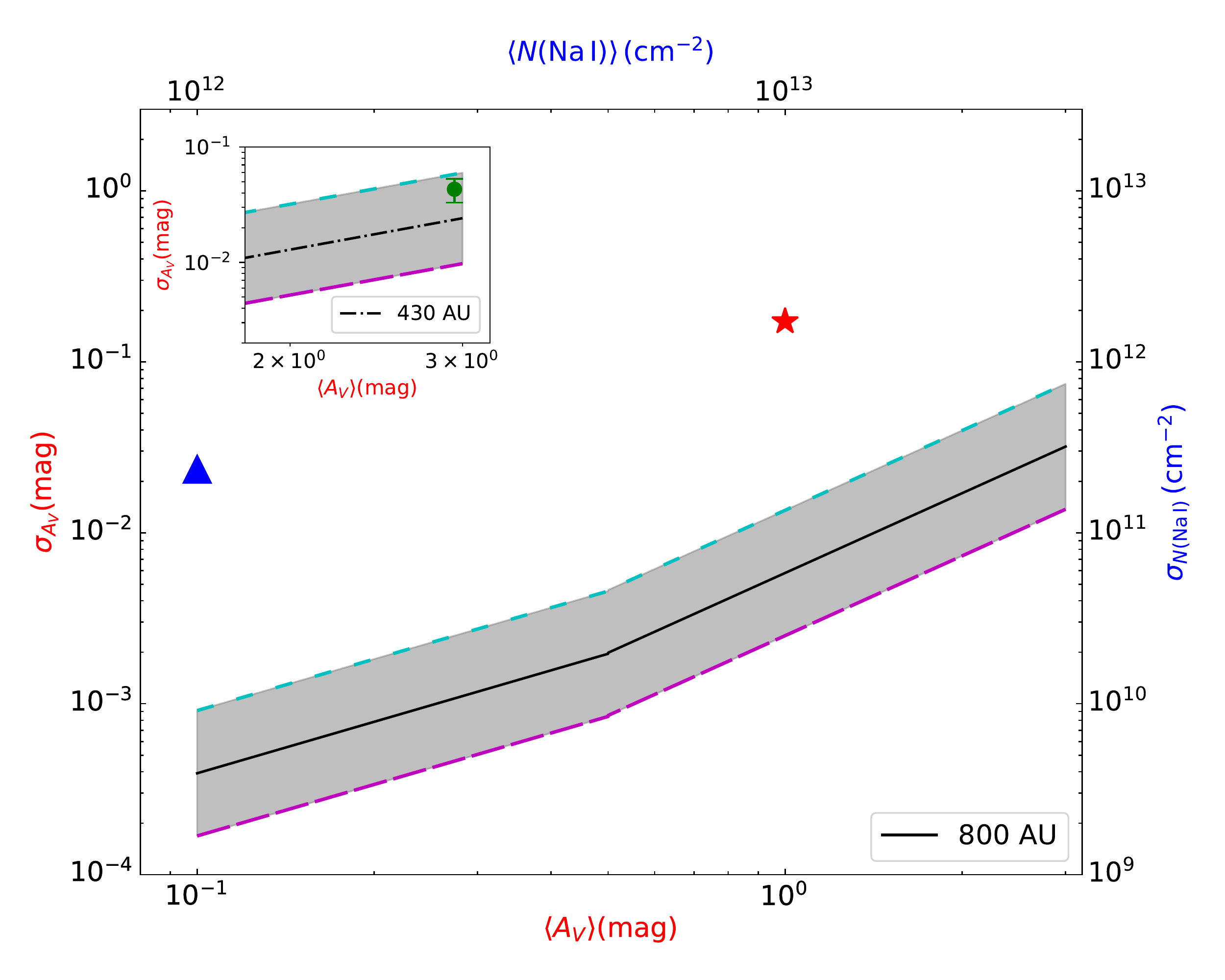}}
\captionof{figure}{{
\ed{The solid line represents the RMS at the 800~AU scale using $\gamma = -2.9 \pm 0.2$ from MD07 (Equation~\ref{eqn:lin-norm-SN-hiAV}), with the cyan (short-dashed) and magenta (long-dashed) lines marking the shallow and steep ends of the $1\, \sigma$ spread, respectively.  The blue triangle indicates the RMS value used in the simulations of P10 (for $N$(\ion{Na}{1})), and the red asterisk, the RMS used in the simulations of F13 (for \AV), respectively, 
translated to the 800~AU scale.  The inset shows the density fluctuations at the 
smaller scale of 430~AU (shifted downward from the RMS for the 800~AU scale by about 25\%), corresponding to the phase of 23.2~days.  The green dot with error bars shows the RMS variation of the observed \AV across phases for \sncu \edd{and therefore will be smaller than the RMS at the latest observed phase shown here.}}}
}\label{fig:AV-Na-sigma}  
\end{minipage}

\subsubsection{Extinction Variations}\label{sec:sn14J-AV}

\citet{brown15a} examined the time variation of the extinction for \snJ.  They found that if they \edd{modeled and applied scattering by circumstellar dust to SN~2011fe, used as the template, the observations of \snJ do not agree with the model predictions.}
\edd{The unavailability of a good twin for \snJ at the present limits quantitative measurements of the \AV variation.}
Here we will just state that, given the observed \AV value of 1.9~mag, 
the \edd{expected upper limits on the RMS} for \AV variation over the phase range of $-10$ to +50 days are: 0.014~mag for $\gamma = -2.9$, and 0.11~mag for $\gamma = -2.5$.

\rff{
We note that light echoes have been observed for \snJ \citep{crotts2015a, yang2017a} at late phases ($> 200$~days).   \citet{bulla2017a} confirmed \rfga{the estimate for the distance between the SN and dust given by} \citet{yang2017a}, and in their analysis concluded the distance is too large for the echoed light to enter the line of sight \rfga{during the range of phases (approximately $-5$ to +25 days) covered by \citet{brown15a}.}
}

In summary, for \sncu and \snJ, both of which are highly reddened, there is little evidence for the time variation of either extinction (\sncu) or EW(\ion{Na}{1}) (\snJ).  This is consistent with the small level of variations \rff{due to non-uniform ISM density} we have predicted. 

\rff{

\rfga{
\subsection{\snX}\label{sec:06X}
SN 2006X is a highly reddened SN Ia that exhibited variable \ion{Na}{1} absorption and variation in the observed \EBV. The variation in \ion{Na}{1} absorption has been attributed either to CSM \citep{patat07a} or a highly specialized ISM configuration \citep{chugai2008a}. Since we predict no detectable variation in \ion{Na}{1} absorption due to random non-uniformity of the \ion{Na}{1} column density across an expanding SN photodisk, we agree that the observed \ion{Na}{1} absorption variation must be due to other causes. The observed change in \EBV over a phase range of $-4$ to +35 days has been interpreted as resulting from a light echo \citep{bulla2017a}, a conclusion that is consistent with the spatially resolved light echo seen from late-time imaging with \HST \citep{wang2008a, crotts2008a}. These analyses agree that the dust causing the light echo has an inferred distance $\gtrsim 20$~pc in the foreground, placing it in the ISM. The observational signatures of non-uniform ISM are quite different from those of a light echo, as we now describe.
}

The scattered photons arriving from a light echo are delayed with respect to the direct light, producing both a brightening and a bluer color. Because the delayed photons arrive from an earlier portion of the light curve, the color and brightness (and the inferred color excess and extinction) become more strongly perturbed after maximum light. 
In addition, the perturbations change smoothly with phase. However, for the case of non-uniform dust, the observed extinction has a random scatter whose amplitude about the mean decreases with time. This occurs because each new epoch has a subset of its photodisk that backlights the same region of ISM that was backlit by the previous epoch. Thus, the extinction distribution across the photodisk at a new epoch will include the (area-averaged) extinction deviations of earlier epochs. As a result, the observable extinction deviations with respect to the (area-averaged) extinction for the largest photodisk observed (i.e., the extinction at the last observed epoch) will be larger for earlier epochs. 

To summarize, the contrast between these phenomena is: 
\begin{itemize}

	\item For a light echo, earlier phases will have values more consistent with each other, while later phases will show brightening and bluer colors.  This will result in large variations in the inferred \EBV and \AV for highly reddened SNe if the dust distance is in the right range \citep{bulla2017a}.
	
	\item For non-uniform ISM dust, the later phases will have values more consistent with each other while earlier phases will show larger and random \rfGA{(but correlated)} variations.  (For example, see Figure~8 for F13: while the variations in F13 are over-stated, as discussed here, the trends are qualitative correct). This effect is independent of the distance between the SN and foreground dust, and we find that typically it will be small even for highly reddened SNe.

\end{itemize}



}

\section{Conclusions}\label{sec:conclusion}


Even though there is clear evidence for ISM density fluctuations on scales 10~AU or even below, we show that due to the steep spectral index of the ISM column density power spectrum and because the SN photodisk is much smaller ($\lesssim 800 \text{ AU}$ in diameter) than the typical ISM structure ($\gtrsim \rm{10\, pc}$), substantial variations on $\gtrsim 10$~pc scales translate to much smaller fluctuations on the scale of SN photodisk.  We provide a way to set the normalization for ISM density fluctuations on length scales relevant to SNe Ia based on MW dust emission observations \citep{miville-deschenes07a}.  
We have found that the expected time variations of EW(\ion{Na}{1}) and \AV for SNe Ia due to non-uniform ISM would be typically much smaller than the predictions in \citet{patat10a} and \citet{foerster13a}, respectively.
The observations of two SNe Ia highly reddened by at least mostly interstellar dust, \sncu and \snJ, are consistent with the results of our analysis.  
\rff{\rfga{The observed color variations for \snX have been attributed to a light echo \citep{bulla2017a} rather than non-uniform ISM, and so are also consistent with our analysis.}}

We therefore conclude, \ed{if non-uniform ISM density in the SN host galaxy follows a scale-free power law power spectrum}: 

1)  The expected variation of extinction for SNe Ia with $\AV < 0.5$~mag is negligible and is certainly  
smaller than what is currently scientifically relevant for SN cosmology.  
This, combined with mounting evidence that interstellar dust is likely the dominant source of reddening for SNe Ia, significantly reduces the concern about time variation of extinction for SNe~Ia \rff{due to non-uniform dust density} in cosmological studies. 

2) \ed{For the study of extragalactic ISM density non-uniformity, it may be possible to detect small variations in extinction across phases for a highly reddened SN.  The RMS can be as large as $\sim 0.1$~mag over a phase range of, say, from $-10$ to +50 days, with the shallowest spectral index ($\gamma = -2.5$).
A larger phase coverage would be better for this purpose (earlier observations would especially help as it allows the observation to sample closer to the full extent of the photodisk at the last observed phase).  \citet{huang17a} demonstrated that in order for such a detection to be feasible, it is important to not only have an unreddened SN that is well-twinned with the dust-obscured SN, but the observed phases of these two SNe need to be closely matched.}  

3) It is very unlikely for the EW of the ISM absorption features to vary with time in any significant way 
\ed{($\sigma_\text{EW} \lesssim 1$~m\AA), even for highly reddened SNe Ia with optimal velocity component distribution. 
Thus any detection of temporal variation in the absorption EW is likely due to other causes.}

\section{Acknowledgement}\label{sec:acknowledgement}
This work was supported in part by the Director, Office of Science, Office of High Energy Physics of the U.S. Department of Energy under Contract No. DE-AC02-05CH11231.  We thank the Gordon \& Betty Moore Foundation for their continuing support.  Additional support was provided by NASA under the Astrophysics Data Analysis Program grant 15-ADAP15-0256 (PI: Aldering).  X.H. acknowledges the University of San Francisco Faculty Development Fund and the support of the Visiting Faculty Program, Office of Science of the U.S. Department of Energy.

\newpage
\appendix
\section*{\RV for Rayleigh Scattering}
\label{sec:appendix}
\ed{In the course of this study, we encountered a comment in \citet{foley14a} concerning} the lowest possible \RV in the limit of Rayleigh scattering.  As is well known, for molecules and small dust particles \citep[$\lesssim 10$~nm for optical wavelengths;][]{kruegel03a}, the scattering cross section depends on the wavelength as $\lambda^{-4}$. 
Contribution from Rayleigh scattering has been invoked as a possible explanation for the sharp blue rise in spectropolarimetry for sight lines toward certain stars \citep{andersson13a} in the MW (scattering by a reflection nebula) and toward SN 2014J \citep{patat15a, hoang17a},
respectively.  In the limit of Rayleigh scattering, \RV would be given by $\RV^{Rayleigh} = (A_B/\AV - 1)^{-1} = (\sigma_B/\sigma_V - 1)^{-1} = ((\lambda_B/\lambda_V)^{-4} - 1)^{-1} 
\approx 0.81$ \citep[e.g.,][]{bessell1990a}.  If the refractive index is complex, in addition to scattering, there is absorption, which goes as $\sim \lambda^{-1}$ \citep{kruegel03a}.  In this case the net power, $p$, for extinction due to small dust particles would be between $-1$ and $-4$.  
For \snJ,  $\RV \approx 1.4$ and the effective power $p$ is between $-2.4$ and $-2.1$ \citep{amanullah14a}, whose absolute value is well below the Rayleigh scattering limit.  Therefore while the \RV value may be extreme for \snJ by MW standards, it is not physically forbidden based on the wavelength dependence of Rayleigh scattering, and \edd{does not rule out pure ISM dust on this basis alone.}

\bibliographystyle{aasjournal}
\bibliography{dustarchive}

\end{document}